\documentclass[runningheads]{llncs}
\usepackage[T1]{fontenc}
\usepackage[table]{xcolor}
\usepackage{graphicx}
\usepackage{amsmath}
\usepackage{booktabs}
\usepackage{overpic}

\definecolor{citecolor}{HTML}{0071BC}
\definecolor{linkcolor}{HTML}{ED1C24}
\usepackage[colorlinks,
            anchorcolor=red,
            citecolor=citecolor, 
            linkcolor=linkcolor,
            ]{hyperref}
            
\newcommand{\figref}[1]{Figure \ref{#1}}
\newcommand{\tabref}[1]{Table \ref{#1}}
\newcommand{\tabincell}[2]{\begin{tabular}{@{}#1@{}}#2\end{tabular}}
\newcommand{\supp}[1]{\textcolor{magenta}{#1}}

\definecolor{ForestGreen}{RGB}{34,139,34}
\definecolor{mygray}{gray}{.92}
\begin{document}
\title{Embracing Massive Medical Data}

\author{Yu-Cheng Chou \and
Zongwei Zhou\thanks{Correspondence to: Zongwei Zhou (\href{mailto:zzhou82@jh.edu}{\texttt{zzhou82@jh.edu}})} \and
Alan Yuille
}
%index{Yu-Cheng Chou}
%index{Zongwei Zhou}
%index{Alan Yuille}
%
\authorrunning{Y.-C. Chou et al.}

\institute{Johns Hopkins University
}

\maketitle

\begin{abstract}

As massive medical data become available with an increasing number of scans, expanding classes, and varying sources, prevalent training paradigms---\textit{where AI is trained with multiple passes over fixed, finite datasets}---face significant challenges. First, training AI all at once on such massive data is impractical as new scans/sources/classes continuously arrive. Second, training AI continuously on new scans/sources/classes can lead to catastrophic forgetting, where AI forgets old data as it learns new data, and vice versa. To address these two challenges, we propose an online learning method that enables training AI from massive medical data. Instead of repeatedly training AI on randomly selected data samples, our method identifies the most significant samples for the current AI model based on their data \textbf{uniqueness} and prediction \textbf{uncertainty}, then trains the AI on these selective data samples. Compared with prevalent training paradigms, our method not only improves data efficiency by enabling training on continual data streams, but also mitigates catastrophic forgetting by selectively training AI on significant data samples that might otherwise be forgotten, outperforming by 15\% in Dice score for multi-organ and tumor segmentation.
The code is available at \href{https://github.com/MrGiovanni/OnlineLearning}{\texttt{https://github.com/MrGiovanni/OnlineLearning}}
\end{abstract}

\section{Introduction}

Massive medical data are becoming publicly available \cite{qu2023abdomenatlas,blankemeier2024merlin}, but are we ready for training AI models on increasing number of scans, varying sources, and expanding classes? Prevalent training paradigms face significant challenges, as training AI all at once on such massive data is impractical~\cite{purushwalkam2022challenges}. 
First, these paradigms are largely designed for fixed, finite datasets. Second, they exhibit signs of catastrophic forgetting when adapted to new scans, sources, or classes~\cite{robins1995catastrophic}.
% Unlike humans, who can retain old knowledge as they acquire new, prevalent training paradigms are limited by their static nature, requiring training on fixed, finite datasets and suffering from catastrophic forgetting when learning new data~\cite{robins1995catastrophic}.

To address these challenges, an ideal training paradigm must meet four requirements: 
\textbf{Req. 1:} handling incomplete annotations, as a combination of public medical datasets is often only partially labeled due to the high cost of annotating medical images at the voxel level \cite{kang2023label}.
Existing methods~\cite{yan2021dynamically,lopez2017gradient} require fully labeled inputs.
\textbf{Req. 2:} learning new scans/sources without forgetting old ones, where prevalent training paradigms that rely on `epoch'---a complete, shuffled pass through the entire dataset---are impractical \cite{zhang2023continual}. Existing methods~\cite{smith2023closer,yan2021dynamically} require AI to see the dataset multiple times.
\textbf{Req. 3:} handling new classes without repeated retraining, as most medical images are not fully labeled yet but will be in the future.
Existing methods~\cite{perkonigg2021dynamic,gonzalez2023lifelong} discard the possibility of new classes. 
\textbf{Req. 4:} training AI with manageable resources, as clinical environments usually lack large storage and computational capacities.
Existing methods~\cite{tiwari2022gcr,ji2023continual} require extensive memory for data selection and class extension.

% \zz{Any well-known papers about online learning, perhaps from OpenAI?}
To meet these requirements, we explore online learning~\cite{lopez2017gradient,chaudhry2018efficient}. 
\textbf{First}, we use text encoding~\cite{liu2023clip} to segment classes based on their text descriptions, allowing us to handle new classes and tasks within partially labeled datasets by simply modifying the text descriptions (Reqs. 1,3).
\textbf{Second}, we store recent samples as new data becomes available, enabling training continuously from new data without revisiting old data and avoiding the need for complete passes through the entire dataset (Req. 2, \S\ref{sec:linear}).
\textbf{Third}, to store the best possible training data, we propose to select samples based on their uniqueness, which prevents catastrophic forgetting by preserving samples from different time periods and sources (Reqs. 2,4, \S\ref{sec:dynamic}).
\textbf{Fourth}, inspired by human learning that focuses on challenging tasks~\cite{anders2008deliberate}, we propose a simple yet effective method that deliberately learns the samples with high uncertainty to improve performance without the need for extra computational resources (Reqs. 4, \S\ref{sec:selective}).

We have examined the proposed method on a large-scale single-site private dataset and a sequential-site dataset composed of $16$ public datasets~\cite{liu2023clip}.
On the single-site dataset, extensive experiments demonstrate that our method achieves comparable performance to the prevalent training paradigm that trains the model over multiple passes (82.2\% vs. 82.6\%).
More importantly, the results in~\tabref{tab:single_score} show that our method enhances data efficiency by enabling training continuously from new data without revisiting old data (\S\ref{sec:data_efficiency}).
On the sequential-site dataset with varying sources of data (\tabref{tab:seq_score}), we observe a 6\% performance increase by effectively selecting unique data compared to directly storing recent samples.
Notably, the results in~\figref{fig:forget} and~\figref{fig:memory}  demonstrate that our method successfully mitigates catastrophic forgetting by enhancing data diversity and preserving previous knowledge (\S\ref{sec:varied}).
Furthermore, by deliberately learning samples with high uncertainty, our method boosts performance by 9\% that even outperforms
the prevalent training paradigm  (\S\ref{sec:deliberate}).

With advancements in automatic annotation \cite{jaus2023towards,wasserthal2023totalsegmentator,li2024well} and the generation of vast synthetic data \cite{hu2023label,chen2024towards,lai2024pixel}, the future of AI seems destined for continuous streams of massive data. Embracing massive medical data, we are among the first to take proactive action that can potentially lead to transformative shifts in AI development. In summary, the main contributions of this work are as follows:

\begin{enumerate}
    % \item To our best knowledge, we are arguably the first to anticipate challenges in handling massive medical data and proactively take action in response. 
    % Our results demonstrate the feasibility of training models on continual streams of massive medical data (\tabref{tab:single_score}).
    \item We show that storing recent samples in the data stream can improve data efficiency, allowing us to train on continual data streams to achieve comparable performance to the prevalent training paradigm (\tabref{tab:single_score}).
    \item We demonstrate the effectiveness of the data deduplication for adapting varying sources, which increases the performance by 6\% on Dice score for multi-organ and tumor segmentation (\tabref{tab:seq_score}).
    \item We follow the human learning pattern to deliberately learn significant samples and further boost performance by 9\%, even outperforming the prevalent training paradigm (\tabref{tab:seq_score}).
    \item We prove the capability of mitigating catastrophic forgetting with both data deduplication and deliberate learning by retaining the previous knowledge (\figref{fig:forget} and \figref{fig:memory}).
\end{enumerate}

\noindent\textbf{\textit{Related Work.}} 
To mitigate catastrophic forgetting, significant research has been conducted in the domain of natural images.
Regularization-based methods constrain model flexibility, typically through techniques applied to weights~\cite{chaudhry2018riemannian} and gradients~\cite{chaudhry2018efficient}, or via knowledge distillation focused on output logits~\cite{castro2018end,schwarz2018progress} and intermediate features~\cite{zhu2022self}. Despite their utility, these methods often fail to ensure optimal performance on challenging tasks, particularly without stored exemplars.
Expandable models address this by either incrementally growing with new tasks~\cite{li2019learn,yoon2017lifelong} or maintaining a common backbone model adaptable through small task-specific adaptation blocks~\cite{mallya2018piggyback}. However, this often leads to inflated model sizes and bloated network architectures.
In contrast, rehearsal techniques involve storing and replaying a small subset of training samples~\cite{buzzega2020dark,liu2021rmm}, embedded features~\cite{iscen2020memory}, or generators~\cite{ostapenko2019learning} from prior tasks. Nonetheless, these methods violate the streaming data setting that our study adheres to.

\section{Method}
\label{sec:method}

% 阐述问题，以及对应的方法段落
To handle continual data streams and dynamically adapt without revisiting old data and forgetting acquired knowledge, we propose 
utilizing a linear memory to store recent samples in the continual data stream (\S\ref{sec:linear}), 
dynamic memory to deduplicate the stored samples (\S\ref{sec:dynamic}), and 
selective memory to learn from the significant samples to further improve the performance (\S\ref{sec:selective}).

\subsection{Linear Memory}\label{sec:linear}
Prevalent training paradigms typically rely on fixed-size datasets, characterized by their finiteness, immutability, and ready availability. 
Leveraging these attributes, samples within such datasets can be conveniently indexed, shuffled, and accessed throughout the training process. 
The prevalent training paradigm often needs iterating over the datasets multiple times. 
In sharp contrast, online learning necessitates a departure from this paradigm, embracing a data stream potentially of infinite length. 
Within this context, data retrieval from a streaming source at any given moment $t$ yields the current sample $x_t$, with future samples inaccessible and past samples retrievable only if stored upon acquisition.

\begin{figure*}[!t]
  \centering
  \includegraphics[width=\linewidth]{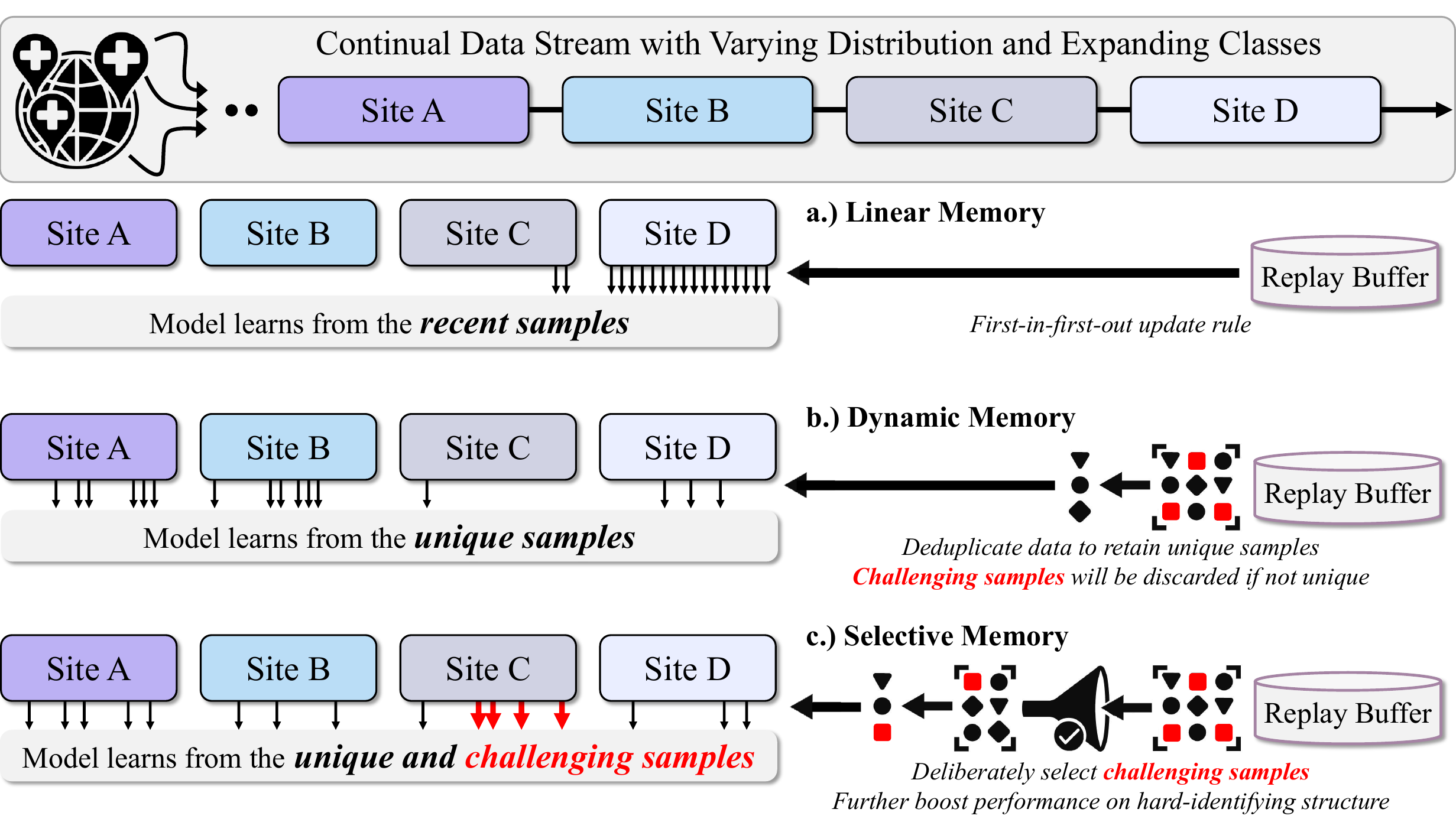}
  \caption{
  \textbf{Different Training Method.}
  Linear memory stores only a few recent samples, causing significant forgetting. Dynamic memory adapts to varying data distributions by retaining unique samples, while selective memory further identifies and selects challenging samples, including those that might be duplicated, ensuring they are not missed by dynamic memory (\S\ref{sec:deliberate}).
  }\label{fig:method}
\end{figure*}

Addressing the challenges, we propose a straightforward solution by leveraging the concept of Linear Memory, tasked with storing a limited number of recent samples. 
This notion draws inspiration from experience replay~\cite{lin1992reinforcement}, a technique widely employed in reinforcement learning~\cite{andrychowicz2017hindsight,mnih2015human,schaul2015prioritized}, supervised continual learning~\cite{hsu2018re,rolnick2019experience}, and self-supervised learning~\cite{purushwalkam2022challenges}.
As shown in \figref{fig:method}-(a), the linear memory serves to store the recent samples from streaming data sources for the training pipeline. 
Specifically, incoming streaming data is seamlessly integrated into the replay buffer $\mathcal{B}$ with memory size $N$, replacing the oldest samples according to a first-in-first-out update rule.
At the same time, mini-batches $x$ of training data can be generated at any time by random sampling from the buffer.
This setup eliminates the need for the training pipeline to revisit previous samples, enabling it to leverage streaming data sources potentially of infinite length.
Simultaneously, linear memory facilitates the reuse of samples through multiple sampling instances, thereby reducing the overall data cost.
As a result, data usage is determined by sampling rate $S$, defined as the ratio between the number of mini-batches and those acquired from the streaming source.

\subsection{Dynamic Memory}\label{sec:dynamic} % dynamic
Medical data obtained from hospitals often exhibit duplicated patterns due to similar sampling strategies or data distribution.
Initially, distributions within a period of time adhere to the same scanning protocol and annotation policy, resulting in a high degree of duplication. 
However, these distributions may shift over time. 
This varying nature stands in sharp contrast to the static data used in the prevalent training paradigm, where models are trained on single, meticulously curated datasets. 
Take AbdomenAtlas dataset~\cite{li2024well} for instance, which allows for random sampling images from a collection of 25 classes. 
However, the continual data stream in online learning setups often challenges this assumption.

To address this challenge, we propose Dynamic Memory, a method designed to optimize the linear memory by retaining only unique samples and actively reducing duplication, as illustrated in \figref{fig:method}-(b).
Consider a replay buffer $\mathcal{B}$ containing $N$ samples, each with a corresponding embedding $z_i$.
When adding a new sample $x_t$ to $\mathcal{B}$, we calculate the cosine similarity between all samples in $\mathcal{B}$. This helps us identify and discard the sample $x_{i^*}$, where $i^*$ is the index in $\mathcal{B}$ with the highest cosine similarity between embeddings $z_i$ and $z_j$ for $i, j \in \mathcal{B}$.
We then employ a moving average to track embedding over iterations.
Note that because each sample represents a small region after data augmentation, rather than the entire input, its embedding shows significant differences. 
By removing duplicated samples from the memory, the correlation between samples decreases, thereby increasing the data diversity in the memory.

\subsection{Selective Memory}\label{sec:selective}

In the realm of varying data streams, we encounter numerous challenges that, intriguingly, parallel the concept of deliberate practice, reminiscent of human learning patterns. 
Different from common practice that simply repetition of certain actions, deliberate practice involves first identifying the problem, and then systematically targeting through specific exercises to correct mistakes~\cite{anders2008deliberate}.
Here, machine learning moves beyond its traditional passive role of processing fixed data, aligning more closely with the active learning mechanisms observed in humans.
Inspired by human learning patterns, we prioritize samples based on their significance, constructing a memory that emphasizes the most challenging samples while discarding the easier ones. 
As a result, our goal is to focus on samples with greater uncertainty and structures carrying heavier penalties.

To achieve this, we introduce Selective Memory (SM).
As shown in~\figref{fig:method}-(c), given a prediction $y$ and the corresponding ground truth $g$, we first calculate the penalty $\alpha_c$ of each structure:

\begin{equation}
\alpha_c =\left\{\begin{array}{l}
         1,~\text{for}~ 0 \textless \hat{\alpha}_c \textless 1 \\
         \hat{\alpha}_c,~ \text{else} \\
    \end{array}\right.
    ,~\text{where}~\hat{\alpha}_c = Norm(\frac{\sum_c^{m}S_c}{S_c})*m,
\end{equation}
and $S_c$ is the size of the corresponding structure, $m$ is the number of structure in $g$ . 
We then apply the obtained penalty $\alpha$ on the entropy calculation $L_{BCE}$ to give the hard-identifying structure more penalty.

To calculate uncertainty $U_x$, we resort to the entropy. 
In information theory, entropy is a measure of uncertainty or unpredictability associated with a random variable~\cite{seidenfeld1986entropy}.
It quantifies the amount of information contained in a message or a set of outcomes, which inherently matches the idea of deliberate practice to identify problems (knowledge with more information) and strategically target them.
As a result, we apply $\alpha$ weighted $L_{BCE}$ to calculate uncertainty $U_x$ and only discard samples from the ones under top $K$\% entropy when calculating the cosine similarity between all pairs of samples:
\begin{equation}
    i,j \in \mathcal{B}_k,~\text{where}~ \mathcal{B}_k=\mathcal{B} \backslash TopK(\{U_x\}).
\end{equation}

\begin{table}[t!]
  \centering
  \scriptsize
  \renewcommand{\arraystretch}{1.15}%{0.95}
  \setlength\tabcolsep{8.5pt}
  \caption{ 
  \textbf{Data Efficiency.}
  The results demonstrate that the linear memory trained on continual data streams achieves comparable performance to the prevalent training paradigm that trains models repeatedly with $100$ epochs.
  Linear memory enables training without the need to revisit old data, thereby enhancing data efficiency.
  Please see Appendix Table \ref{tab:supp_jhh_score} for full results.
  }
  \begin{tabular}{l||ccc|c} 
  \toprule
  \multicolumn{5}{c}{\tabincell{c}{\textbf{Single-Site Dataset}}} \\
  \hline
  Strategy &Linear Memory &Linear Memory &Linear Memory &Repeatedly\\ % e90
  \hline
  Epochs &- &- &- &100 \\
  Sampling Rate $S$ &100 &100 &100 &- \\
  Memory Size $N$ &64 &128 &256 &- \\
  \hline
  Average Dice &0.8217 &0.8222 &0.8225 &0.8260\\
  \bottomrule
  \end{tabular}
  \label{tab:single_score}
\end{table}

\section{Experiment \& Result}\label{sec:experiment_result}

\subsection{Experimental Setting}\label{sec:details}
\noindent\textbf{\textit{Dataset.}}
We adopt two large-scale CT datasets in our experiments, including a single-site private dataset \cite{park2020annotated,xia2022felix} and a sequential-site dataset~\cite{liu2023clip}.
(1) Our \textbf{single-site} dataset comprises abdominal CT scans with per-pixel annotation, encompassing $15$ classes of abdominal organs, all sourced from a single hospital with a consistent distribution.
We split the dataset randomly into $2,101$ training cases and $516$ testing cases.
(2) Our \textbf{sequential-site} dataset consists of $16$ partially labeled sub-datasets ($D_d, d \in \left[1,16\right]$), collectively offering $32$ classes.
This combination of sub-datasets incorporates various annotation policies and scanning protocols, mirroring the varying distributions of a data stream.
We adhered to the data splitting protocol in~\cite{liu2023clip}, resulting in $2,100$ training cases and $583$ testing cases.
%, 
It is worth noting that we did not shuffle the data in any of our experiments for the data stream setting.

\noindent\textbf{\textit{Implementation and baseline.}}
For adapting partially labeled sub-datasets, we adopt the CLIP-Driven Universal Model~\cite{liu2023clip} for multi-organ and tumor segmentation.
U-Net~\cite{ronneberger2015u} is chosen as the backbone, and we follow the default training setting in~\cite{liu2023clip} with the standard binary cross-entropy loss and the Dice loss.
We adopt the Dice score as the metric to evaluate the results in our experiments.
Note that the prevalent training paradigm trains models repeatedly with $100$ epochs, which violates the streaming setting in our experiments.

\begin{table}[t!]
  \centering
  \scriptsize
  \renewcommand{\arraystretch}{1.15}%{0.95}
  \setlength\tabcolsep{8.5pt}
  \caption{ 
  \textbf{Dynamic Adaptation.}
  Under the varying distributions in the streaming source, Dynamic Memory (DM) and Selective Memory (SM) enable the identification of the significant samples and thereby enhance the segmentation performance.
  Please see~\supp{supp. material} for full results. 
    }
  \begin{tabular}{l||ccccc|c} 
  \toprule
  \multicolumn{7}{c}{\tabincell{c}{\textbf{Sequential-Site Dataset~\cite{liu2023clip}}}} \\
  \hline
  Strategy &LM &DM &SM &SM &SM &Repeatedly\\ % e90
  \hline
  Epochs &- &- &- &- &- &100 \\
  Sampling Rate $S$ &100 &100 &100 &100 &100 &- \\
  Memory Size $N$ &128 &128 &128 &128 &128 &- \\
  Top K\% &- &- &12.5\% &25\% &50\% &- \\
  \hline
  Tumor Average Dice &0.2559 &0.3125 &0.3436 &\textbf{0.3878} &0.3426 &0.3520 \\
  Organ Average Dice &0.3471 &0.4066 &0.4805 &\textbf{0.4958} &0.4834 &0.4783 \\
  \hline
  Average Dice       &0.3272 &0.3860 &0.4505 &\textbf{0.4722} &0.4525 &0.4506 \\
  \bottomrule
  \end{tabular}
  \label{tab:seq_score}
\end{table}

\begin{figure}[!t]
  \centering
  \includegraphics[width=\linewidth]{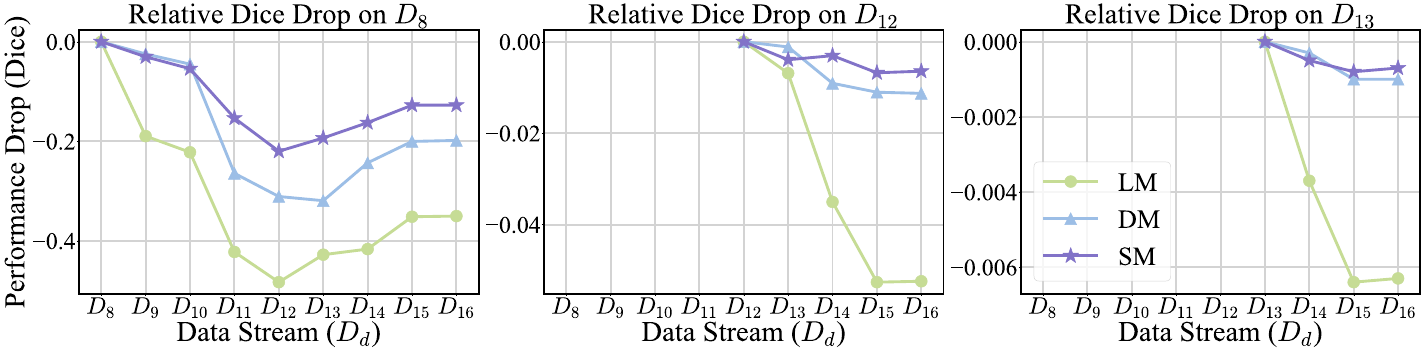}
  \caption{
  \textbf{Catastrophic Forgetting.} 
  To evaluate forgetting, we calculate the relative Dice drop after training on the incoming sub-datasets. 
  Both DM and SM store samples from previous sub-datasets, thereby alleviating forgetting observed with LM.
  }\label{fig:forget}
\end{figure}

\subsection{Linear Memory Enables Training Without Revisiting Old Data}\label{sec:data_efficiency}
We investigate the effectiveness of linear memory by training on the continual data stream.
As shown in~\tabref{tab:single_score}, we trained the model with and without linear memory on our single-site dataset with a consistent distribution.
Linear memory notably enhances data efficiency in the continual data stream setting without the need to revisit old data. 
Additionally, under the same amount of updates, the linear memory with a sampling rate $S = 100$ achieved similar performance to the prevalent training paradigm.
\tabref{tab:single_score} also shows that by increasing memory size $N$, the performance raised slightly due to increasing samples in the memory.
However, this enhancement comes with a notable memory overhead.
Consequently, we set $N=128$ for a cost-accuracy trade-off.

\subsection{Dynamic Memory Mitigates Catastrophic Forgetting}\label{sec:varied}
Compared to our single-site dataset, the sequential-site dataset~\cite{liu2023clip} features varying distributions in scanning protocol, annotation policy, and slicing scope.
This variability leads to significant performance degradation, as illustrated in~\tabref{tab:seq_score}.
While the naive linear memory can somewhat mitigate this issue, it fails to match the performance of the prevalent training paradigm (32.7\% vs. 45.1\%) and still experiences substantial forgetting.
Specifically, linear memory shows a marked performance decline on $D_d$ when learning from the incoming sub-datasets, as shown in~\figref{fig:forget}.
Conversely, the dynamic memory, which retains unique samples from each sub-dataset, demonstrates considerable improvements (\figref{fig:memory}). 
This method boosts performance by 6\% (\tabref{tab:seq_score}) and effectively mitigates catastrophic forgetting, as evidenced in~\figref{fig:forget}.

\subsection{Selective Memory Outperforms Prevalent Training Paradigms}\label{sec:deliberate}
\begin{figure}[!t]
  \centering
  \includegraphics[width=\linewidth]{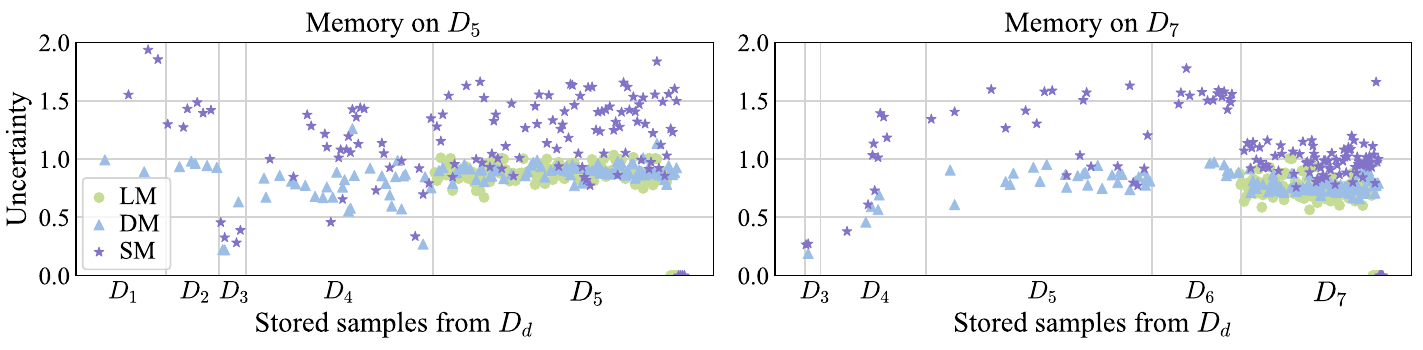}
  \caption{
  \textbf{Diverse Memory.} 
  We visualize the memory to demonstrate the diversity of stored samples from previous $D_d$.
  Both DM and SM can retain the samples from previous sub-datasets.
  SM can further identify samples with higher uncertainty.
  }\label{fig:memory}
\end{figure}

Although dynamic memory can retain unique samples, the performance remains unsatisfactory.
By deliberately targeting the samples with higher uncertainty, selective memory significantly increases the performance that even outperforms the prevalent training paradigm (47.2\% vs. 45.1\%, \tabref{tab:seq_score}) and is able to identify small structures such as Adrenal Gland (detailed in~\supp{supp. material}).
Especially, as shown in~\figref{fig:memory} (right), compared to dynamic memory, selective memory stores the high uncertainty samples of $D_6$ when training on $D_7$.
This illustrates that selective memory can identify challenging but potentially duplicated samples that are ignored by dynamic memory.
Last, we ablate this method with different $K\%$ settings and set $K=25\%$ to ensure a balance of diversity and uncertainty that maximizes the model's performance.

\section{Discussion \& Conclusion}\label{sec:discussion_conclusion}

As the volume of public medical data continues to expand, prevalent training paradigms struggle to handle the continual streams of massive medical data seamlessly. 
Our study, as the pioneer in anticipating this trend and proactively taking action in response, introduced linear memory and dynamic memory to address data inefficiencies while retaining significant samples to effectively accommodate varying distributions.
Additionally, inspired by human learning patterns, we introduced selective memory to further actively focus on challenging samples.
We experimentally show that our method not only surpasses the prevalent training paradigm but also significantly mitigates catastrophic forgetting.
In conclusion, we present potential solutions tailored to evolving clinical scenarios, and we believe that further exploration in this direction is required to continually learn by more fine-grained and, even, no annotation.

\begin{credits}
\subsubsection{\ackname} This work was supported by the Lustgarten Foundation for Pancreatic Cancer Research and the Patrick J. McGovern Foundation Award.

\subsubsection{\discintname}
The authors have no competing interests to declare that are relevant to the content of this article.
\end{credits}

% ---- Bibliography ----
\bibliographystyle{splncs04}
\bibliography{Paper-0065}

\appendix
\newpage

\begin{table}[t!]
  \centering
  \scriptsize
  \renewcommand{\arraystretch}{1.15}%{0.95}
  \setlength\tabcolsep{8.5pt}
  \caption{ 
  \textbf{Dynamic Adaptation.}
  Under the varying distribution in the streaming source, the proposed Dynamic Memory (DM) and Selective Memory (SM) enable the identification of the critical samples and thereby enhance the segmentation performance.
  Specifically, SM significantly outperforms other methods on complex structures such as the Esophagus and small structures such as the Adrenal Gland.
    }
  \begin{tabular}{l||cc|ccc|c} 
  \toprule
  \multicolumn{7}{c}{\tabincell{c}{\textbf{Sequential-Site Dataset~\cite{liu2023clip}}}} \\
  \hline
  Strategy &LM &DM &SM &SM &SM &Prevalent\\ % e90
  \hline
  Sampling Rate $S$ &100 &100 &100 &100 &100 &- \\
  Memory Size $N$ &128 &128 &128 &128 &128 &- \\
  Top K\% &- &- &12.5\% &25\% &50\% &- \\
  \hline
  Spleen &0.9506 &0.9543 &0.9553 &0.9449 &0.9542 &0.9268 \\
  Right Kidney &0.8961 &0.9236 &0.9239 &0.9202 &0.9122 &0.9189 \\
  Left Kidney &0.8972 &0.9172 &0.9145 &0.9138 &0.9052 &0.9149 \\
  Gall Bladder &0.2205 &0.6747 &0.5367 &0.5906 &0.6325 &0.3671 \\
  \rowcolor{mygray}
  Esophagus &0.0004 &0.1588 &0.4376 &0.4044 &0.4537 &0.0395 \\
  Liver &0.9615 &0.9673 &0.9668 &0.9665 &0.9632 &0.9630 \\
  Stomach &0.5802 &0.7705 &0.7313 &0.7890 &0.7467 &0.7785 \\
  Aorta &0.4895 &0.6063 &0.6751 &0.5687 &0.6633 &0.7676 \\
  Postcava &0.2098 &0.5373 &0.6191 &0.5444 &0.5239 &0.6433 \\
  Vein &0.0000 &0.0370 &0.0000 &0.0000 &0.0000 &0.2310 \\
  Pancreas &0.6669 &0.8028 &0.8249 &0.8112 &0.7766 &0.7376 \\
  \rowcolor{mygray}
  Right Adrenal Gland &0.0000 &0.0000 &0.5830 &0.5942 &0.4307 &0.0000 \\
  \rowcolor{mygray}
  Left Adrenal Gland &0.0000 &0.0000 &0.5625 &0.5126 &0.4716 &0.0000 \\
  Duodenum &0.1204 &0.4293 &0.4000 &0.4987 &0.5251 &0.3599 \\
  Hepatic Vessel &0.4678 &0.5571 &0.5769 &0.5524 &0.5731 &0.5063 \\
  Right Lung &0.7002 &0.4919 &0.5432 &0.6112 &0.5529 &0.7617 \\
  Left Lung &0.8771 &0.6187 &0.7965 &0.7162 &0.7859 &0.9102 \\
  Colon &0.0008 &0.0003 &0.0185 &0.0352 &0.0187 &0.4986 \\
  Intestine &0.0000 &0.0849 &0.1081 &0.2765 &0.3648 &0.4610 \\
  Rectum &0.0000 &0.0000 &0.0000 &0.0000 &0.0102 &0.0000 \\
  Bladder &0.6383 &0.5397 &0.7316 &0.7701 &0.7722 &0.7517 \\
  \rowcolor{mygray}
  Prostate &0.0009 &0.0923 &0.1066 &0.2276 &0.0471 &0.0000 \\
  Left Head of Femur &0.0000 &0.0000 &0.0000 &0.0772 &0.0000 &0.0000 \\
  Right Head of Femur &0.0000 &0.0000 &0.0000 &0.0690 &0.0000 &0.4190 \\
  Celiac Truck &0.0000 &0.0000 &0.0000 &0.0000 &0.0000 &0.0000 \\
  \hline
  Kidney Tumor         &0.2417 &0.2936 &0.2483 &0.3258 &0.3204 &0.3781 \\
  Kidney Cyst          &0.0268 &0.0059 &0.0729 &0.2920 &0.0056 &0.1751 \\
  Liver Tumor          &0.3579 &0.5753 &0.5755 &0.6059 &0.5817 &0.6219 \\
  Pancreas Tumor       &0.1643 &0.2823 &0.2957 &0.3435 &0.3560 &0.2058 \\
  Hepatic Vessel Tumor &0.5717 &0.6659 &0.6516 &0.6618 &0.6713 &0.5930 \\
  Lung Tumor           &0.1910 &0.1845 &0.2862 &0.1981 &0.1516 &0.3233 \\
  Colon Tumor          &0.2377 &0.1797 &0.2751 &0.2874 &0.3106 &0.1667 \\
  \hline
  Tumor Average Dice       &0.2559 &0.3125 &0.3436 &\textbf{0.3878} &0.3426 &0.3520 \\
  Organ Average Dice        &0.3471 &0.4066 &0.4805 &\textbf{0.4958} &0.4834 &0.4783 \\
  \hline
  Average Dice              &0.3272 &0.3860 &0.4505 &\textbf{0.4722} &0.4525 &0.4506 \\
  \bottomrule
  \end{tabular}
  \label{tab:supp_comp_score}
\end{table}

\begin{table}[t!]
  \centering
  \scriptsize
  \renewcommand{\arraystretch}{1.15}%{0.95}
  \setlength\tabcolsep{8pt}
  \caption{ 
  \textbf{Data Efficiency.}
  By integrating linear memory into the prevalent training paradigm, we enable training on continual data streams without the need to revisit old data, thereby enhancing data efficiency.
  The results demonstrate that the linear memory trained on continual data streams achieves comparable performance to the prevalent training paradigm.
  }
  \begin{tabular}{l||ccc|c} 
  \toprule
  \multicolumn{5}{c}{\tabincell{c}{\textbf{Proprietary Dataset \cite{xia2022felix,park2020annotated}}}} \\
  \hline
  Strategy & Linear Memory & Linear Memory & Linear Memory & Repeatedly \\ % e90
  \hline
  Sampling Rate $S$ &100 &100 &100 &- \\
  Memory Size $N$ &64 &128 &256 &- \\
  \hline
  Aorta               &0.8865 &0.8893 &0.8890 &0.8848 \\
  R Adrenal Gland &0.7530 &0.7534 &0.7501 &0.7491 \\
  L Adrenal Gland  &0.7001 &0.6980 &0.6962 &0.6964 \\
  Celiac Truck        &0.5403 &0.5428 &0.5385 &0.5610 \\
  Colon               &0.6915 &0.6946 &0.6992 &0.7196 \\
  Duodenum            &0.7907 &0.7903 &0.7878 &0.7896 \\
  Gall Bladder        &0.8866 &0.8888 &0.8860 &0.8886 \\
  Postcava            &0.8129 &0.8088 &0.8161 &0.8164 \\
  Right Kidney        &0.9535 &0.9536 &0.9551 &0.9527 \\
  Left Kidney         &0.9473 &0.9477 &0.9475 &0.9460 \\
  Liver               &0.9715 &0.9712 &0.9717 &0.9708 \\
  Pancreas            &0.8699 &0.8683 &0.8703 &0.8688 \\
  Intestine           &0.6047 &0.6088 &0.6144 &0.6375 \\
  Spleen              &0.9664 &0.9661 &0.9663 &0.9641 \\
  Stomach             &0.9512 &0.9509 &0.9506 &0.9453 \\
  \hline
  Average Dice &0.8217 &0.8222 &0.8225 &0.8260\\
  \bottomrule
  \end{tabular}
  \label{tab:supp_jhh_score}
\end{table}

\end{document}